\documentclass[aps,pra,twocolumn,superscriptaddress,linenumbers]{revtex4-2}
\usepackage[T1]{fontenc}
\usepackage{soul}
\usepackage{textcomp}
\usepackage[english]{babel}
\usepackage{graphicx}
\usepackage{gensymb} 
\usepackage{dcolumn}
\usepackage{bm}
\usepackage{bm}
\usepackage{amsmath}
\usepackage{verbatim}     
\usepackage[dvipsnames]{xcolor}       
\usepackage[normalem]{ulem}   
\usepackage{bbold}
\usepackage{mathrsfs}
\usepackage[mathscr]{eucal}
\usepackage{physics}      
\newcommand{\qo}[1]{``#1''}                               		
\newcommand{\beq}{\begin{equation}}
\newcommand{\eeq}{\end{equation}}
\newcommand{\bei}{\begin{itemize}}			
\newcommand{\eei}{\end{itemize}}			

\usepackage{hyperref}
\hypersetup{
   pdfpagemode=None,
   pdfstartpage=1,
   pdfmenubar=true,
   pdftoolbar=true,
   colorlinks = true,
   linkcolor=blue,
   citecolor=blue,
   urlcolor=blue,
   bookmarksopen=false
 }

\definecolor{lgreen}{RGB}{15,150,15}

\begin{document}
\nolinenumbers

\title{Compact and programmable large-scale optical processor in free space}

\author{Maria Gorizia Ammendola}
\affiliation{Scuola Superiore Meridionale, Via Mezzocannone, 4, 80138 Napoli, Italy}
\affiliation{Nexus for Quantum Technologies, University of Ottawa, K1N 5N6, Ottawa, ON, Canada}
\affiliation{Dipartimento di Fisica, Universit\`{a} degli Studi di Napoli Federico II, Complesso Universitario di Monte Sant'Angelo, Via Cintia, 80126 Napoli, Italy}

\author{Nazanin Dehghan}
\affiliation{Nexus for Quantum Technologies, University of Ottawa, K1N 5N6, Ottawa, ON, Canada}
\affiliation{National Research Council of Canada, 100 Sussex Drive, Ottawa ON Canada, K1A 0R6}

\author{Lukas Scarfe}
\affiliation{Nexus for Quantum Technologies, University of Ottawa, K1N 5N6, Ottawa, ON, Canada}

\author{Alessio D'Errico}
\affiliation{Nexus for Quantum Technologies, University of Ottawa, K1N 5N6, Ottawa, ON, Canada}
\affiliation{National Research Council of Canada, 100 Sussex Drive, Ottawa ON Canada, K1A 0R6}


\author{Francesco Di Colandrea}\email{francesco.dicolandrea@unina.it}
\affiliation{Nexus for Quantum Technologies, University of Ottawa, K1N 5N6, Ottawa, ON, Canada}
\affiliation{Dipartimento di Fisica, Universit\`{a} degli Studi di Napoli Federico II, Complesso Universitario di Monte Sant'Angelo, Via Cintia, 80126 Napoli, Italy}
\author{Ebrahim Karimi}
\affiliation{Nexus for Quantum Technologies, University of Ottawa, K1N 5N6, Ottawa, ON, Canada}
\affiliation{National Research Council of Canada, 100 Sussex Drive, Ottawa ON Canada, K1A 0R6}
\affiliation{Institute for Quantum Studies, Chapman University, Orange, California 92866, USA}

\author{Filippo Cardano}\email{filippo.cardano2@unina.it}
\affiliation{Dipartimento di Fisica, Universit\`{a} degli Studi di Napoli Federico II, Complesso Universitario di Monte Sant'Angelo, Via Cintia, 80126 Napoli, Italy}

\begin{abstract}
\noindent
Photonic circuits are central to classical and quantum information processing. While integrated technologies dominate, free-space architectures are emerging as attractive alternatives, offering broad bandwidth and direct manipulation of optical modes without confinement in waveguides. A key challenge for scalability lies in circuit depth, as the number of layers manipulating the optical field typically grows with the system size. Here, we introduce a programmable free-space photonic platform that performs high-dimensional unitary transformations using only three layers. Information is encoded in structured light modes defined by circular polarization and quantized transverse momenta, and processed with spatial light modulators interleaved with half-wave plates. We implement unitaries that are equivalent to quantum walks over up to 30 time steps, in one- and two-dimensional lattices, distributing a single input mode across more than 7,000 outputs, where conventional approaches would require tens or hundreds of layers. Despite being restricted to translationally-invariant systems, the platform supports diverse quantum walk dynamics, including disorder, synthetic gauge fields, and topological effects, previously explored only in separate experiments. Using coincidence detection with a time-tagging camera, we show compatibility with quantum optics protocols and provide examples of quantum walks of heralded single photons. These results contribute to establish free-space optical processors as promising resources for high-dimensional quantum simulation and scalable optical information processing.
\end{abstract}

\maketitle
\section{Introduction}
Programmable photonic platforms are versatile tools for classical and quantum technologies, enabling applications in communication~\cite{Luo2023}, information processing~\cite{Slussarenko2019}, and simulation~\cite{Flamini2018}. They are considered essential building blocks for all-optical quantum computers~\cite{Wang2025} and photonic neural networks~\cite{Fu2024}, thanks to their ability to manipulate spatial, temporal, spectral, and polarization degrees of freedom. Integrated circuits exploit waveguide arrays with beam splitters and phase shifters to couple spatial modes~\cite{wang2020integrated,Peruzzo2025}, while temporal modes are controlled with integrated optics~\cite{morandotti25}, fiber loops~\cite{Schreiber2012}, or birefringent materials for ultrafast operations~\cite{Bouchard2024}. Alternatively, multimode fibres combined with spatial light modulators (SLMs) give access to large Hilbert spaces of transverse modes, supporting high-dimensional entanglement~\cite{Goel2240}, quantum walks~\cite{Makowski2024}, and reconfigurable quantum networks~\cite{valencia2025multiplexed,Leedumrongwatthanakun2020}.\\
Free-space optical processors offer a compelling alternative to integrated solutions, providing flexible access to many co-propagating structured modes. By combining programmable phase masks with free-space propagation \cite{Labroille2014}, they have attracted interest for optical computing and photonic neural networks~\cite{Hu2024,Chen2025}. Examples of achievable tasks are mode sorting, high-dimensional quantum gates, entanglement certification, and photonic quantum computing~\cite{Brandt2020,Kupianskyi2023,Defienne2023,Koni2024,Wang2024a,Lib2024}. However, their scalability is limited by circuit depth and optical losses. Here we address this challenge by introducing a reconfigurable free-space platform based on liquid-crystal SLMs and an analytical inverse-design method, enabling exact unitary transformations with only three patterned layers.
Diffractive elements implementing space-variant polarization manipulation~\cite{Rubin2021b} have further expanded the toolbox for free-space processors. Besides high-dimensional quantum gates~\cite{Nagali2009a,Stav2018}, recent works realized quantum walks of structured light~\cite{Cardano2015,Derrico20}, underscoring their potential for scalable quantum information processing. Yet, in most architectures, the circuit depth scales with the number of modes, making low-loss, accurate, and programmable implementations a central challenge.\\
To mitigate this, recent studies have focused on resource-efficient few-layer platforms. Single-layer dielectric metasurfaces have implemented generalized beam splitters~\cite{Yousef2025}, C-NOT gates~\cite{Li2025}, and multiphoton state characterization~\cite{An2024}. Complementarily, by using multi-layer patterned liquid-crystal devices, an efficient compression scheme for implementing large-scale unitaries in one- and two-dimensional spatial configurations using only three patterned layers has been demonstrated~\cite{DiColandrea2023,ammendola2024large}.\\
Building on this strategy, here we integrate these compression techniques into reconfigurable circuits realized with commercially available LC-based SLMs. Unlike static LC devices, SLMs provide dynamic programmability via computer control. Our circuit-design method yields analytical solutions for translation-invariant unitary transformations, avoiding iterative optimization approaches typically used in multi-plane light converters~\cite{Brandt2020,Kupianskyi2023,Lib2024} or propagation through complex media~\cite{Leedumrongwatthanakun2020,Goel2240}. This allows, in principle, exact realization of arbitrary unitaries with only three devices.\\
Although SLMs are usually employed as phase-only elements on scalar fields, their use for space-dependent polarization transformations has been proposed with cascaded configurations~\cite{Sit2017,HU2020125028,Hu_2021} and partially demonstrated with a single device~\cite{Sit2017}. While this work was in preparation, an experimental realization of such transformations for structured light was reported in a related setup~\cite{Forbes25}. In combination with our results, these advances establish the first complete reconfigurable platform for arbitrary, space-dependent polarization transformations.\\
We validate the platform by implementing over 300 distinct mode-coupling unitaries in the form of coined quantum walks~\cite{VenegasAndraca2012}, while maintaining fixed and shallow circuit depth. We further demonstrate its suitability for quantum optics by integrating event-based, single-photon-sensitive cameras that simultaneously resolve transverse positions and nanosecond photon arrival times~\cite{Nomerotski2019}. Together, these results represent a significant step toward stable, efficient, and scalable free-space photonic circuits.

\section{Results}
\subsection{Coupling transverse momentum modes via SLMs}
\begin{figure*}
    \includegraphics[width=\linewidth,  trim = 0.25cm 4cm 0.25cm 0.25cm, clip
    ]{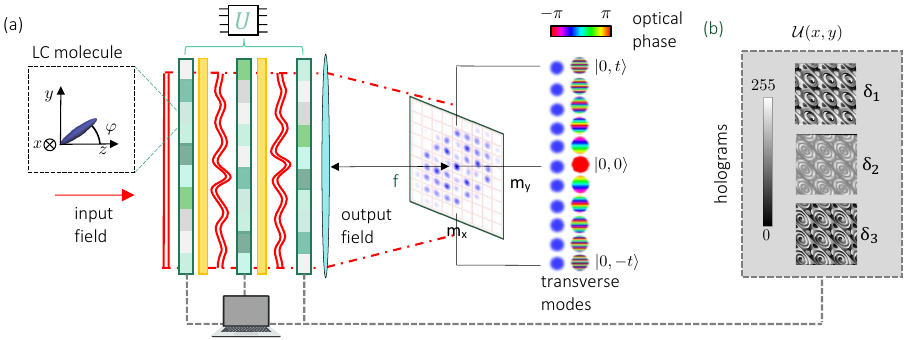}
    \caption{\textbf{Programmable photonic circuit based on spatial light modulators.} (a) Conceptual overview of the platform. The input field $\ket{0,0,j}$ propagates through three SLMs, alternating with two HWPs (yellow), implementing the optical transformation corresponding to the target circuit. The output field modal distribution is revealed in the focal plane of a lens, where the optical intensity pattern is recorded on a camera. Photonic modes at the circuit input and output carry ($m_x,m_y$) units of transverse momentum $\Delta k_{\perp}$ along the $x$ and $y$ directions, respectively, resulting in linear phase gradients across the transverse $xy$ plane. For simplicity, a 1D cut along ${m_x=0}$ is shown at the output plane. Optical polarization provides an additional degree of freedom, effectively doubling the dimensionality of the mode space. The inset shows the local orientation of the LC molecules within the SLM. The local birefringent retardation $\delta(x,y)$ depends on the local out-of-plane orientation $\varphi$. 
     (b)~A dedicated software computes the holograms for each SLM and displays them on three SLMs, enabling real-time reprogramming of the circuit for specific target unitaries.
    }
    \label{fig:fig1}
\end{figure*}

Our circuit processes optical modes having the following expression:
\begin{equation}\label{eqn:opticalmodes}
    \ket{m_x,m_y,j}= A(x,y,z) e^{i k_z z} e^{i (m_x x+m_y y) \Delta k_{\perp}}\ket{j},
\end{equation}
where $(x,y)$ are the coordinates in the transverse plane, with photons assumed to propagate along $z$. 
Here, $A(x,y,z)$ is a Gaussian envelope with beam waist $w_0$, $k_z$ is the wavevector $z$-component, $\Delta k_{\perp}$ is a unit of transverse momentum, $\ket{j}$ is a polarization state, which can be left-handed $\ket{L}=(1, 0)^T$ or right-handed $\ket{R}=(0, 1)^T$, and $(m_x,m_y)$ are integer numbers.

A conceptual scheme of the setup is sketched in Fig.~\ref{fig:fig1}(a). Starting from a single spatial mode, e.g.\ ${\ket{0,0}}$, the application of a unitary map $U$, invariant under discrete translations in the Hilbert space spanned by $\ket{m_x,m_y}$, results in a superposition of output modes:
\begin{equation}\label{eqn:modemixing}
    U \ket{0,0,j}=\sum^t_{m'_x,m'_y=-t} \sum_{h \in \lbrace L,R\rbrace}c_{m'_x,m'_y,h}\ket{m'_x,m'_y,h},
\end{equation}
where $t$ sets the range of transverse modes coupled by $U$. Consequently, when considering a localized input state, the total number of addressable transverse modes via $U$ is $d^2 = (2t+1)^2$. Measuring the optical field in the focal plane of a lens, where the modes defined in Eq.~\eqref{eqn:opticalmodes} are spatially resolved, enables rapid and direct retrieval of the field distribution across the circuit’s optical modes, without requiring additional mode projection or scanning. Importantly, negligible overlap between neighbouring modes is guaranteed as long as $w_0\geq\Lambda$~\cite{Derrico20}, where $w_0$ is the beam waist.

Our photonic circuit implements the target unitary $U$ as a coined QW, a widely used model for discrete-time quantum dynamics on a lattice with an internal degree of freedom, referred to as the coin. The QW describes a sequence of unitary operations that combine coin rotations with coin-dependent translations across lattice sites (more details in Sec.\ \ref{subsubsec:1d2dqw}). In our implementation, lattice sites correspond to spatial modes $\ket{m_x,m_y}$ and coin states to circular polarizations $\ket{j=L,R}$~\cite{Derrico20}.
This encoding enables a compact implementation of translationally-invariant unitary maps via space-dependent polarization transformations~\cite{DiColandrea2023, ammendola2024large} (see Methods for details).
In this framework, the target unitary operator can be also modelled as a position-dependent polarization transformation acting on the structured light field, and takes the form:
\begin{equation}
U=\iint \text{d}x\,\text{d}y\,\, \mathcal{U}(x,y) \otimes\ketbra{x,y},
\label{eqn:digonal}
\end{equation}
where $\mathcal{U}(x,y)$ is an arbitrary SU(2) matrix, defined by three real parameters. The translation invariance of $U$ implies that $\mathcal{U}(x,y)$ is periodic with period ${\Lambda=2\pi/\Delta k_{\perp}}$, and contains spatial frequency components corresponding to quantized values of the transverse momentum unit $\pm \Delta k_{\perp}$~\cite{DiColandrea2023}. Put simply, the effect of this polarization transformation can be viewed as polarization-controlled diffraction, imparting momentum kicks to photons in discrete steps of $\Delta k_{\perp}$, which matches the spacing between neighbouring modes in momentum space (see Eq.~\eqref{eqn:opticalmodes}).

Previous studies have shown that $\mathcal{U}(x,y)$ can be realized via a minimal set of three waveplates with patterned optic-axis orientations, implementable using LCMSs~\cite{DiColandrea2023,ammendola2024large}. Each metasurface acts as a standard waveplate with a spatially varying optic-axis, $\theta(x,y)$, given by the in-plane orientation of LC molecules with respect to the $x$ axis, and uniform yet tunable birefringence $\delta$. By applying an electric field to the plate, LC molecules are tilted out-of-plane toward the propagation direction, which allows for controlling the phase difference between the ordinary and extraordinary components, as shown in the inset of Fig.~\ref{fig:fig1}(a). The metasurfaces' patterns $\theta_i(x,y)$ (${i=\lbrace1,2,3\rbrace}$) realizing the transformation of Eq.~\eqref{eqn:digonal} are then found by imposing
\begin{equation}
\mathcal{U}(x,y)=Q_{\theta_3(x,y)}(\pi/2)Q_{\theta_2(x,y)}(\pi)Q_{\theta_1(x,y)}(\pi/2),
\end{equation}
where
\begin{equation}
Q_{\theta}(\delta)= 
    \begin{pmatrix}
    \cos(\frac{\delta}{2}) &&i\sin(\frac{\delta}{2}) e^{-2 i \theta}\\
    i\sin(\frac{\delta}{2}) e^{2 i \theta} && \cos(\frac{\delta}{2})
    \end{pmatrix},
    \label{eqn:jonesmatrix}
\end{equation}
is the standard waveplate Jones matrix in the circular polarization basis. 
%

However, such a circuit is a static machine capable of implementing only the target operation. To achieve reconfigurability, we coherently replaced LCMSs with SLMs. 
Essentially, an SLM consists of a pixelated array of LC-filled cells, each of which can be controlled independently via software~\cite{Forbes:16, Yang2023}. In the circular polarization basis, its Jones matrix can be written as
\begin{equation}\label{eqn:slmjones}
S_{\theta}(\delta(x,y))=e^{i \frac{\delta(x,y)}{2}}Q_{\theta}(\delta(x,y)),
\end{equation}
where, opposite to LCMSs, $\delta$ can be locally controlled by applying a different electric voltage at each pixel, while the in-plane orientation $\theta$ is uniform. In the following, we assume ${\theta=0}$.
SLMs are traditionally employed as digital phase retarders inducing a phase shift ${\delta (x,y)}$ to the incoming $\ket{H}$-polarized beam, where ${\ket{H}=\left(\ket{L}+\ket{R}\right)/\sqrt{2}}$ is the horizontal polarization, that is parallel to the SLM optic-axis.
The voltage, and hence the phase shift, can be controlled using grayscale images, referred to as holograms, usually with an $8$-bit encoding, which allows for $256$ phase levels. 
Here, we exploit the possibility of programming the local birefringence parameter of SLMs in order to use them as polarization-controlling devices, effectively acting as waveplates with space-dependent optical retardation. 
At each transverse position, the target unitary $\mathcal{U}(x,y)$ can be implemented via three SLMs, locally programmed to satisfy the equation
\begin{equation}\label{eqn:solve}
   \mathcal{U}(x,y)= e^{-i \frac{\delta_0}{2}} S_{0}(\delta_3)H_2S_{0}(\delta_2)H_1S_{0}(\delta_1),
\end{equation}
where ${\delta_0=\delta_1+\delta_2+\delta_3}$, ${H_1=Q_{\pi/8}(\pi)}$ and ${H_2=Q_{-\pi/8}(\pi)}$ are two half-wave plates (HWPs), and we omitted the dependence of each $\delta_i$ (${i=\lbrace1,2,3\rbrace}$) on $(x,y)$ for ease of notation. The intermediate HWPs are necessary to prevent the trivial effect of three cascaded SLMs, which would result in a mere phase transformation on the $\ket{H}$ polarization component.
Moreover, to account for the spatially varying global phase $-\delta_0/2$, we always set $\ket{H}$ at the input of the circuit, and superimpose the phase mask on the first (phase-only) hologram:
\begin{equation}
S_0(\delta_1)\rightarrow S_0(\tilde{\delta}_1),
\end{equation}
with ${\tilde{\delta}_1=\delta_1-\delta_0/2}$.
To simulate the output of a different polarization input, say $U\ket{\phi}$, with ${\ket{\phi}=\Omega\ket{H}}$, we implement the rotated operation
\begin{equation}
\mathcal{U'}(x,y)\ket{\phi}= S_{0}(\delta_3)H_2S_{0}(\delta_2)H_1S_{0}(\Tilde{\delta}_1) \ket{H},
\end{equation}
where ${\mathcal{U'}=\Omega^{-1}\mathcal{U}\Omega}$. 
This phase compensation and input polarization rotation could also be implemented using a fourth SLM placed at the input of the platform, although here we aim to keep the circuit depth as low as possible.\\
The analytical solutions of Eq.~\eqref{eqn:solve}, yielding the set of holograms for a given unitary $U$, are provided in Methods. Via software, these solutions are uploaded and displayed on the SLMs (see Fig.~\ref{fig:fig1}(b)). 
To eliminate the undesired effect of free-space propagation between consecutive SLMs, we interpose two $4$-$f$ systems between them. A complete description of the experimental setup is also provided in Methods. 

\subsection{Experimental Results}
We realize multiple steps of different QW processes across 1D and 2D lattices (Sec.~\ref{subsubsec:1d2dqw}), also observing the effects of time-dependent step operators on the spreading of the wavefunction (Sec.~\ref{subsubsec:superdiff}-~\ref{subsubsec:bloch}).
Additionally, we show how the platform can be engineered to simulate either localized input states (Sec.~\ref{subsubsec:WN}) or wavepackets (Sec.~\ref{subsubsec:QM}), without making any changes in the experimental setup, thus allowing to probe geometrical and topological features in chiral-symmetric processes.
Finally, we test the platform in the single-photon regime, (Sec.~\ref{subsubsec:singlephoton}), validating its suitability for quantum experiments.

\subsubsection{Time-resolved 1D and 2D QWs}\label{subsubsec:1d2dqw}
As anticipated, in our encoding, the walker lattice is spanned by the transverse modes introduced in Eq.~\eqref{eqn:opticalmodes}, while circular polarization states encode a 2-level coin. 
Each run of the experiment corresponds to a fixed time step $t$ of the QW dynamics. In this implementation, the SLMs display the three holograms $\delta_i(x,y)$ corresponding to the unitary $U(t)=U_0^t$, where $U_0$ is the single-step QW evolution operator and $t$ is the number of time-steps. 
Specifically, we focus on the QWs introduced in Ref.~\cite{Cardano2015} and Ref.~\cite{Derrico20} for the 1D and 2D configuration, respectively, where the single-step evolution operators are
\begin{equation}\label{eqn:protocols}
\begin{split}
U_{1}(\alpha_1)&=MT_x(\alpha_1)WM^{\dagger},\\
U_{2}(\alpha_2)&=T_y(\alpha_2)T_x(\alpha_2)W.
\end{split}
\end{equation}
 Here, $W$ is the coin rotation operator, reading
\begin{equation}
W=\frac{1}{\sqrt{2}}
\begin{pmatrix}
    1 && i\\
    i && 1
\end{pmatrix},
\end{equation}
while
\begin{equation}
M=
\begin{pmatrix}
    \cos (\pi/8) && i\sin(\pi/8)\\
    i\sin(\pi/8) && \cos (\pi/8)
\end{pmatrix},
\end{equation}
and
\begin{equation}\label{eq:tx}
\begin{aligned}
&T_x(\alpha)=
\cos(\alpha/2) \mathbb{I}_{c} \otimes \mathbb{I}_{w} + i\sin(\alpha/2)\\
& \sum_{m_x}\ketbra{R}{L}\otimes \hat{t}^\dagger_x +\ketbra{L}{R}\otimes \hat{t}_x
\end{aligned}
\end{equation}
is the coin-dependent translation operator along $m_x$, where ${\hat{t}_x \ket{m_x}=\ket{m_x-1}}$, and ${\mathbb{I}_{c} \otimes \mathbb{I}_{w}}$ is the identity operator acting on the coin and walker space. A similar expression holds for $T_y(\alpha)$. The parameter $\alpha$ tunes the hopping amplitudes between neighboring sites. In the experiment, we set $\alpha_1=\pi$ and $\alpha_2=\pi/2$. 

\begin{figure*}\includegraphics[width=\linewidth,  trim = 0.2cm 4cm 0.25cm 0.25cm, clip]{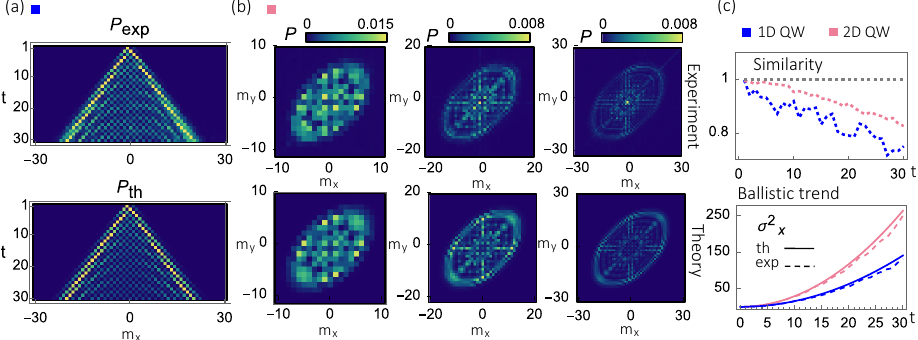}
    \caption{\textbf{One- and two-dimensional quantum walks implemented via three SLMs.}
    (a)~Experimental $P_{\text{exp}}$ (top) and theoretical $P_{\text{th}}$ (bottom) probability distributions for $30$ time steps of the 1D QW protocol $U_1$, with input state $\ket{L}$. (b)~Experimental (top) and theoretical (bottom) probability distributions $P$ for $30$ time steps of the 2D QW protocol $U_2$, with input state $\ket{H}$. Representative distributions are shown for time steps that are multiples of 5. The full time-resolved 2D dynamics is available in the Supplementary Material. (c)~Similarity (top) and corresponding trend of the variance of the output distribution (bottom) for the 1D (blue) and the 2D (pink) QWs. The experimental data points (dots) correctly reproduce the expected ballistic behaviour (solid lines) extracted numerically.} 
    \label{fig:fig2}
\end{figure*}

The input is a localized walker state, obtained by preparing a Gaussian beam with beam waist ${w_0/\Lambda\geq1}$~\cite{Derrico20}. The source is an 810~nm diode laser coupled to single-mode fiber for spatial filtering (see Methods).
At the output of the circuit, the QW distribution can be resolved in the focal plane of a lens placed after the last SLM. Each light spot is associated with a walker site, with probability given by the integrated light intensity within that spot, normalized with respect to the total light power. The procedure to simultaneously control the three SLMs and extract the output probability distribution in real-time is outlined in Methods.
The agreement between the theoretical predictions and the experimental observations is quantified in terms of the similarity
\begin{equation}\label{similarity}
S=\left(\sum_{m_x,m_y}\sqrt{P_\text{{exp}}(m_x,m_y) P_\text{{th}}(m_x,m_y)}\right)^2,  
\end{equation}
where $P_\text{{exp}}$ and $P_\text{{th}}$ are the normalized experimental and theoretical probability distributions, respectively.
Figure~\ref{fig:fig2}(a-b) shows the experimental distributions obtained for the 1D and 2D protocols described above, for input states $\ket{0,L}$ and $\ket{0,0,H}$, respectively, compared with theoretical predictions. We report the simulations of up to 30 steps in both realizations. In the 2D case, this corresponds to a QW activating up to $2*(2*30+1)^2\simeq7400$ modes, where the factor 2 accounts for the polarization degree of freedom. Figure~\ref{fig:fig2}(a) shows the complete step-by-step 1D dynamics, while only multiples of 5 steps are reported in Fig.~\ref{fig:fig2}(b) for the 2D case.
The full time-resolved 2D dynamics can be found in the Supplementary Material. This step-by-step analysis showcases the remarkable advantage provided by reconfigurability compared to the previous LCMS-based static platform, which would have required fabricating a different set of three plates for each target simulation.
The decrease in similarity values with higher step numbers, shown in the top panel in Fig.~\ref{fig:fig2}(c), can be ascribed to the reduced resolution and larger pixel size of the last SLM used in this experiment compared to the first two, leading to aliasing effects. This results in an increased amount of light which remains coupled to the origin for higher step numbers.

A key feature of the QW is that its probability distribution spreads faster than the classical random walk. In particular, the variance of the distribution scales quadratically with the number of steps, $\sigma^2\propto t^2$, which makes the QW a ballistic process. 
The bottom panel in Fig.~\ref{fig:fig2}(c) shows the variances extracted experimentally at each step in the two realizations. In the 2D case, the variance along the $x$ direction is chosen for reference.

\subsubsection{Superdiffusive transition induced by temporal disorder}\label{subsubsec:superdiff}
In the previous section, we assumed the single-step operator to be identical at each step, though this condition can be relaxed to obtain more general transformations associated with time-dependent QWs. These processes can be employed to simulate temporal disorder and external fields, as we show in the following.

\begin{figure*}
    \includegraphics[width=\linewidth,  trim = 0.2cm 4cm 0.25cm 0.25cm, clip]{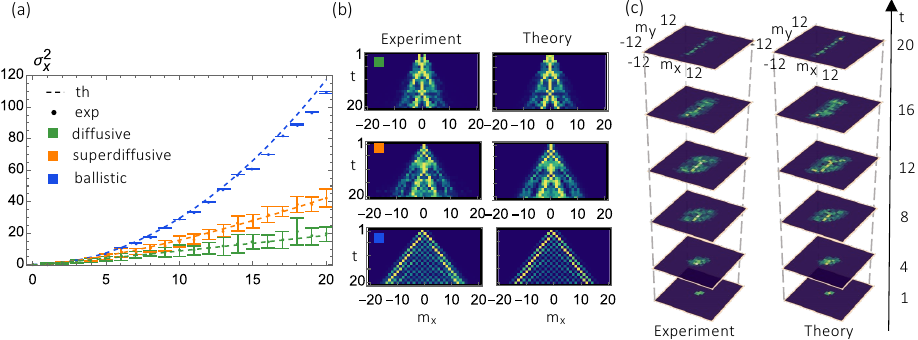}
    \caption{\textbf{Dynamical disorder and electric quantum walks.}
    (a)~Theoretical (dashed) and experimental (solid) trend of the variance of the walker distributions in diffusive (green), superdiffusive (orange), and ballistic (blue) regimes. The experimental values of the variance are averaged over $5$ disorder realizations.
    (b)~Experimental (left) and theoretical (right) output probability distributions for a 1D QW with temporal disorder in a representative diffusive (top), superdiffusive (middle), and ballistic (bottom) configuration. For each regime, the remaining $4$ disorder configurations can be found in the Supplementary Material. (c)~Experimental (left) and theoretical (right) probability distributions showing directional refocusing induced by a constant electric field in a 2D QW ($U_2$). Representative distributions are shown for time steps that are multiples of $4$. The full time-resolved 2D dynamics is available in the Supplementary Material.}
    \label{fig:fig3}
\end{figure*}

By introducing temporal disorder in a QW, it is possible to modify its variance behavior, $\sigma^2(t)\propto t^{\beta}$~\cite{VenegasAndraca2012}.
By varying the disorder strength, we experimentally investigated the transition from ballistic (${\beta=2}$) to diffusive (${\beta=1}$), which is a clear transition from a quantum-to-classical stochastic process, also accessing the intermediate superdiffusive regime (${1<\beta<2}$).
Disorder is introduced by simulating a time-dependent translation operator:
\begin{equation}
U(t)=\prod^t_{n=1}U_1(n)=\prod^t_{n=1} T_x(\alpha_1(n))W,
\end{equation}
with ${\alpha_1(n)}$ randomly extracted from the range ${\left[\bar{\alpha}_1-\Delta \pi,\bar{\alpha}_1+ \Delta \pi\right]}$, where ${\bar{\alpha}_1=\pi/2}$ and $\Delta$ gives the degree of disorder~\cite{Geraldi19}.
Experimental data in Fig.~\ref{fig:fig3}(a) are in excellent agreement with the theoretical prediction (dashed lines) for the three chosen disorder strengths. Each datapoint is obtained as the average over $5$ different realizations within each disorder regime, with error bars given by the corresponding standard deviations. The input is a localized $\ket{0,H}$ state.
Table~\ref{tab:betas} reports the chosen values of $\Delta$ and the values
of ${\beta}$ obtained by fitting the experimental data ($\beta_{\text{exp}}$), compared to the values obtained through numerical simulations ($\beta_{\text{th}}$) of the same disordered evolutions. Uncertainties on ${\beta_\text{exp}}$ are derived from a weighted fit.
The programmability of our platform enables a time-resolved investigation of this transition encompassing up to 30 steps, surpassing previous experimental demonstrations~\cite{Geraldi19}. Figure~\ref{fig:fig3}(b) shows experimental distributions for one value of $\Delta$ for each disorder strength. All the remaining data can be found in the Supplementary Material.
\begin{table}[h!]
\centering
\begin{tabular}{lccc}
\hline
 & $\Delta$ & $\beta_{\text{th}}$ & $\beta_{\text{exp}}$ \\ 
\hline
\text{ballistic} & $0\%$ & $1.99$ & $1.92 \pm 0.05$ \\ 
\text{superdiffusive} & $37.5\%$ & $1.5$ & $1.4 \pm 0.1$ \\
\text{diffusive} & $87.5\%$ & $1.2$ & $1.2 \pm 0.3$ \\
\hline
\end{tabular}
\caption{ Expected theoretical ($\beta_{\text{th}}$) and experimental ($\beta_{\text{exp}}$) values of the exponent $\beta$ for increasing disorder strengths ($\Delta$), inducing the transition from ballistic to superdiffusive to diffusive spreading.}
\label{tab:betas}
\end{table}

\subsubsection{Bloch oscillations in an electric 2D QW}\label{subsubsec:bloch}
When an external force is applied to the walker, the QW can be used to simulate the effect of an electric field on a charged particle. Here, we reveal this effect in a 2D electric walk.
As shown in previous works~\cite{Derrico20,Derrico20bloch}, the application of an external constant force $F_x$ is equivalent to a transverse displacement of the single-step evolution operator according to:
\begin{equation}
    \mathcal{U}(x,y,t)=\prod^t_{n=0}\mathcal{U}_0(x + \Delta x(n),y),
\end{equation}
where ${\Delta x(n)= n F_x \Lambda/ (2 \pi)}$. In our experiment, we set ${U_0=U_2(\pi/2)}$ (see Eq.~\eqref{eqn:protocols}). The energy bands resulting from this protocol exhibit an energy gap ${E_g \approx 1}$~\cite{Derrico20}, which is sufficiently larger than the applied force ${F_x=\pi/10}$, thereby ensuring the validity of the adiabatic approximation. 
Under this approximation, an input state localized in the position space $\ket{m_x, m_y}$ refocuses in the direction of the applied force with the typical period of Bloch oscillations ${T=2 \pi/F_x= 20}$ time steps~\cite{Chalabi2020}. This effect is well captured by our experimental simulation of $20$ steps of the evolution of a localized $\ket{0,0,H}$ input, as shown in Fig.~\ref{fig:fig3}(c), where only multiples of 4 steps are reported. The full time-resolved dynamics can be found in the Supplementary Material.
\subsubsection{Measurement of the winding number in a topological QW}\label{subsubsec:WN}
\begin{figure*}
    \includegraphics[width=\linewidth,  trim = 0.25cm 2.2cm 0.25cm 0.25cm, clip]{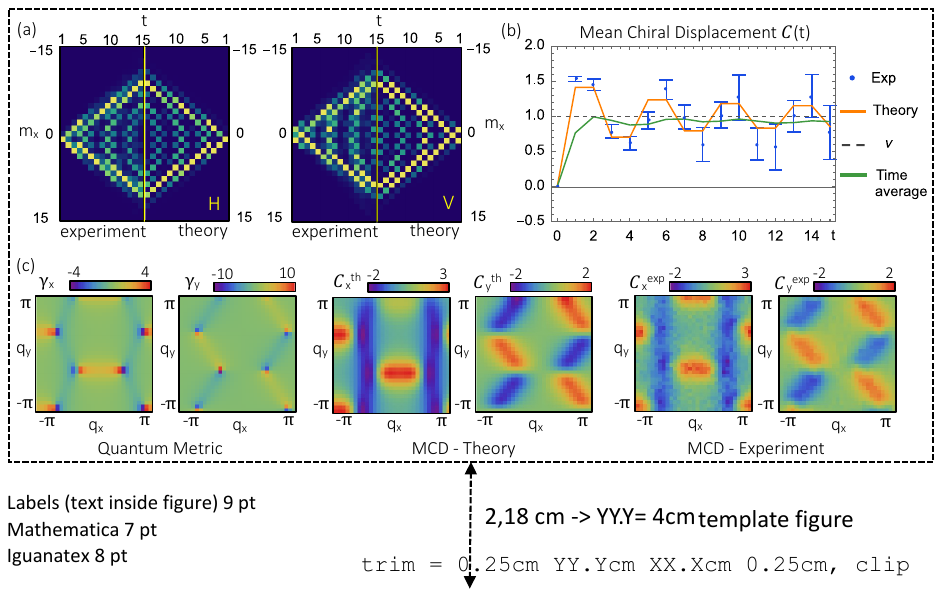}
    \caption{\textbf{Measurement of topological features in one- and two-dimensional lattices.}
    (a)~Experimental reconstruction (left) and theoretical prediction (right) of the output distributions corresponding to the two chiral polarization components. (b)~Experimental reconstruction (blue dots) and theoretical prediction (orange line) of the MCD asymptotically converging to the winding number $\nu=1$ (grey dashed line). The green line shows the time average of the MCD. Error bars are extracted from Monte Carlo simulations, assuming a maximum uncertainty of ${\pm\pi/10}$ on the local birefringence of the three SLMs over 10 different simulated experimental runs.
    (c)~Quantum metric elements $\gamma_{x}$ and $\gamma_{y}$ (left), theoretical prediction (center) and experimental reconstruction (right) of the MCD components $C_x$ and $C_y$ for a $21 \times 21$ grid of $\vec{q}$-values in the Brillouin zone.}
    \label{fig:fig4}
\end{figure*}

QWs provide a paradigmatic example of a periodically driven (Floquet) system that can be engineered to host all topological phases of single-particle systems~\cite{kitagawa2010}.
These are characterized by quantized global features, known as topological invariants, underlying remarkable physical phenomena, such as robust edge states and quantized transport~\cite{Hashemi2025}.
For instance, the 1D QW protocol $U_1$ defined in Eq.~\eqref{eqn:protocols} exhibits chiral symmetry, characterized by the existence of a unitary operator ${\Gamma}$ that pairs states with opposite energy. This symmetry implies the quantization of the Zak phase: ${\varphi_\text{Z}=\nu\pi/2}$, where $\nu$ is an integer, called the winding number, playing the role of a symmetry-protected topological invariant~\cite{zak1989berry,kitagawa2010, asboth2016short}.
By choosing the single-step QW operator to be ${U_0=\tilde{U}_1=\Tilde{R} U_1(\pi) \tilde{R}^{\dagger}}$, with $\tilde{R}=(\mathbb{I}_{c}-i\sigma_y)/\sqrt{2}$, the chiral operator is ${\Gamma= \sigma_x}$.
The mean chiral displacement (MCD) is an observable introduced in Ref.~\cite{Cardano2017}, defined as ${\mathcal{C}_x(t):=2\langle{\Gamma x}\rangle}$. It can be experimentally retrieved by measuring the weighted difference between the center of mass of the intensity distributions of the two chiral projections, i.e.,
the projections on the eigenstates of the chiral operator, that are ${\ket{\uparrow}=\ket{H}}$ and ${\ket{\downarrow}=\ket{V}}$ for the protocol considered above.
When considering a walker that is initialized on a single lattice site with an arbitrary coin state,  $\ket{\psi(0)}=\ket{0,\phi}$, the MCD asymptotically converges to the winding number $\nu$~\cite{Cardano2017}:
\begin{equation}\label{eq:mcd}
    \mathcal{C}_x(t)= 2\sum_{m_x} m_x (P_{\uparrow}(m_x,t)-P_{\downarrow}(m_x,t))\xrightarrow{t \to \infty} \nu.
\end{equation}
By monitoring the evolutions of the two chiral projections (see Fig.~\ref{fig:fig4}(a)), we successfully reconstructed the MCD convergence to ${\nu =1}$, as expected for this QW protocol (see Fig.~\ref{fig:fig4}(b)). Our simulation refers to the evolution of a localized $\ket{H,0}$ input across 15 time steps, thus surpassing the previous realizations of Ref.~\cite{Cardano2017} (7 steps) and~\cite{Derrico20} (13 steps).
Error bars are extracted from Monte Carlo simulations, performed by assuming a maximum uncertainty of ${\pm\pi/10}$ on the local birefringence of the three SLMs over 10 simulated experimental runs.

\subsubsection{Measurement of the quantum metric in a 2D lattice}\label{subsubsec:QM}
In our setup, the walker lattice sites are mapped into optical modes carrying $(m_x,m_y)$ units of transverse momentum ${2\pi/\Lambda}$. Localized states correspond to Gaussian beams with ${\omega_0/\Lambda\geq 1}$~\cite{Derrico20}, which are focused into separated spots in the Fourier plane. As a consequence, the walker quasi-momentum $\vec{q}$ 
maps into the transverse position ${(x,y)}$ on the SLMs' plane (see Methods).
Within this encoding, the hologram periodicity $\Lambda$ corresponds to one Brillouin Zone (BZ), with $\vec{q}\in$ BZ $=[-\pi,\pi]^2$.
Accordingly, preparing a narrow wavepacket peaked around ${\vec{q}_0}$,
\begin{equation}
    \ket{\psi(0)}=\int_{\text{BZ}} \frac{\text{d}^2q}{(2\pi)^2} G_{\omega_0,\vec{q}_0}(\vec{q}) \ket{\vec{q},\phi_0},
\end{equation}
where ${G_{\omega_0,\vec{q}_0}= \mathcal{N} \exp(-(\vec{q}-\vec{q}_0)^2/\omega_0^2)}$, with $\mathcal{N}$ a normalization factor, corresponds to preparing a beam having waist ${w_0/\Lambda \ll 1}$ in the $(x,y)$ plane, centered in ${(x_0,y_0)=(q_{0x},q_{0y})\Lambda/2\pi}$, having polarization $\ket{\phi_0}$.
Our approach allows for the preparation of these wavepackets without making any changes to the setup. Instead of preparing a smaller beam waist as done in previous works~\cite{Derrico20, qmDiColandrea24}, we expand the periodicity of the holograms displayed on the SLMs so that the beam is still covering the same region of each SLM, but this now corresponds to only a portion of the BZ. Then, we digitally shift the holograms so that the center of the beam matches ${\vec{q}_0}$. In so doing, our platform grants parallel access to the simulation of localized (delocalized) wavefunctions in the quasi-momentum space by simply \qo{zooming-in} (\qo{zooming-out}) the computed holograms.
In our experiment, we specifically target the evolution operator generated by a 2D graphene-like chiral-symmetric Hamiltonian, having chiral operator
$\Gamma=\sigma_z$. 
The eigenstructure of this model is detailed in the Supplementary Material. 
In Ref.~\cite{qmDiColandrea24}, some of us showed that the MCD of wavepackets sharply peaked in the quasi-momentum space of tight-binding models featuring chiral symmetry is directly related to the elements of the quantum metric $\gamma$~\cite{PhysRevLett.121.020401,Gianfrate2020}, whose components are defined as $\gamma_i= (\mathbf{n}(\vec{q})\times \partial_{q_i}\mathbf{n}(\vec{q}))\cdot\hat{\bm{\textbf{z}}}$, where $\textbf{n}$ is the pseudo-spin Bloch eigenstate~\cite{kitagawa2010}.
The MCD components along the $x$ and $y$ directions for an input $\ket{\psi(0)}$ are~\cite{qmDiColandrea24}
\begin{equation}\label{eqn:qm-mcd}
    \mathcal{C}_i(t) = 2 \int_{BZ} \frac{\text{d}^2q}{(2\pi)^2} \abs{G_{\omega_0,\vec{q}_0}(\vec{q})}^2 \sin^2{(t E(\vec{q}))} \gamma_i(\vec{q}),
\end{equation}
where ${i=\lbrace x,y\rbrace}$.
We set $t=1$ and compute the three holograms corresponding to the evolution obtained from the graphene Hamiltonian with flat bands:
\begin{equation}
    \mathcal{U}_g(x,y)\equiv\mathcal{U}_g(\vec{q})= e^{- i \Bar{E} \textbf{n}(\vec{q}) \cdot \hat{\bm{\sigma}}},
\end{equation}
where ${\hat{\bm{\sigma}}=\left(\sigma_x,\sigma_y,\sigma_z\right)}$ is the vector of the three Pauli matrices, and we choose ${\Bar{E}=\pi/2}$, so that the oscillating factor ${\sin^2{(\Bar{E}t)}}$ is equal to 1. 
This allows us to experimentally reconstruct the convolution between the Gaussian wavepacket and the quantum metric via a single MCD measurement.
In particular, we magnify the BZ so that $\Tilde{\Lambda}=7 \Lambda$ (see Supplementary Material for further details). We shift the three holograms to center the beam in $21 \times 21$ different $\vec{q}_0$ values. Then we measure the difference between the average $m_x$ and $m_y$ positions of the two output chiral projections, applying the lattice size scaling  $\Delta \Tilde{k}_{\perp}= 2 \pi /\Tilde{\Lambda}$ and comparing it with theoretical predictions (see Fig.~\ref{fig:fig4}(c)) obtained from the simulation of the complete wavepacket dynamics via Eq.~\eqref{eqn:qm-mcd}.
Interestingly, in the proximity of the Dirac points, the quantum metric diverges while the MCD collapses to zero~\cite{qmDiColandrea24}. This can be understood by observing that a Gaussian wavepacket initialized in the vicinity of a Dirac point develops a ring-shaped distribution due to the spin-orbit coupling ${\mathbf{q}\cdot \boldsymbol{\sigma}}$ induced by the local Hamiltonian in the low-energy limit. This solid-state effect is analog to the optical spin-orbit coupling induced by $q$-plates \cite{marrucci2006optical}.

\subsubsection{Heralded single-photon experiment}\label{subsubsec:singlephoton}
To assess the suitability of the circuit for quantum experiments, we simulate different single-particle dynamics by employing a heralded single-photon source. Temporally correlated photon pairs are generated via Spontaneous Parametric Down-Conversion (SPDC), then separated and coupled into single-mode fibers. One channel is used as a herald, while the other photon undergoes the QW. The SPDC source is well described in Methods. Both photons are subsequently directed onto an event-based camera (TPX3CAM) for coincidence measurements, as shown in Methods. Featuring a nanosecond time resolution~\cite{Nomerotski2019}, the TPX3CAM enables the extraction of spatially resolved coincidences without the need for an external trigger. This comes at the cost of lower detection efficiency. 
By analysing a few selected pixels corresponding to the herald and the QW region, we isolate simultaneously detected photons, which allows us to reconstruct the probability distribution resulting from the QW.

\begin{figure*}
    \includegraphics[width=\linewidth,  trim = 0.25cm 2.0cm 0.25cm 0.25cm, clip]{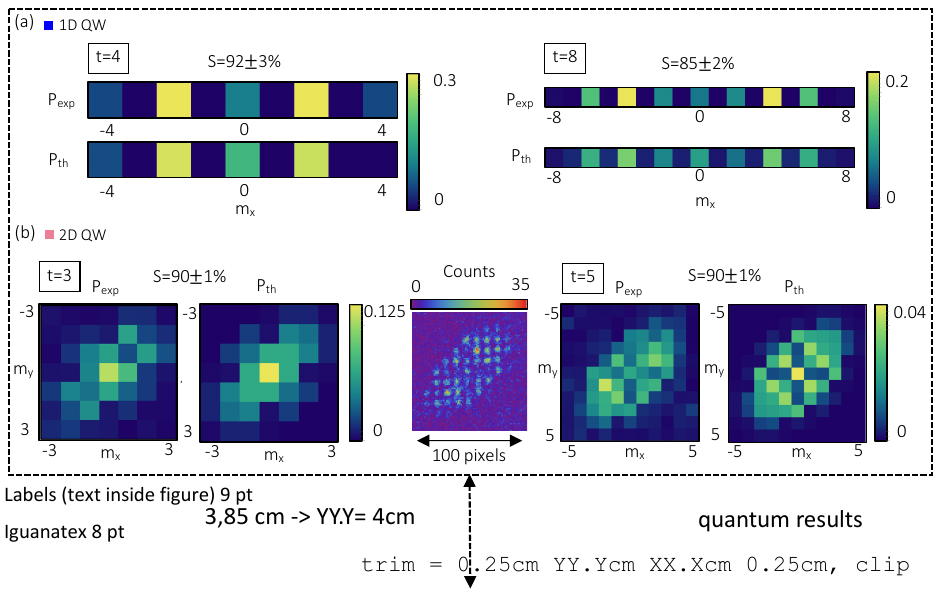}
    \caption{\textbf{Single-photon experiment.}
    Experimentally reconstructed ($\text{P}_{\text{exp}}$) and theoretically predicted ($\text{P}_{\text{th}}$) probability distributions after (a) $4$ and $8$ time steps in 1D, and (b) $3$ and $5$ time steps in 2D quantum walks. Statistical uncertainties on the similarity ($S$) are computed assuming Poissonian statistics for photon counts. The central inset in panel (b) shows the output after $5$ steps on the TPX3CAM ($256 \times 256$ pixels, pixel size $55.5\,\mu$m), following a $30$-minute acquisition and post-selection of coincidence events.}
    \label{fig:fig5}
\end{figure*}

Figure~\ref{fig:fig5} shows the extracted 1D and 2D QW distributions. Panel (a) compares experimental and theoretical results for the 1D QW, while panel (b) shows the corresponding 2D distributions using the heralded single-photon source. The central inset in panel (b) displays the TPX3CAM output after post-selecting coincidence events. Errors on the similarity (S) are extracted from Poissonian photon count statistics.

The resulting single-photon distributions closely resemble those of the laser-driven QW, mainly differing in the noise level. This is expected, as our data contain a relatively high-background photon count. Although time correlations help suppress this background, a residual component persists. Further background reduction can be achieved by selecting events with a time delay away from the temporal correlation peak (where only background counts are recorded) and subtracting this offset from the coincidences. This approach enhances the signal-to-noise ratio by minimizing the background noise in the distributions.

\section{Discussion}
We have demonstrated a reconfigurable photonic circuit implementing a wide class of unitary transformations, via optical manipulation at three layers only, corresponding to commercial SLMs.
Its performance has been validated both with a classical laser source and single photons, implementing more than $300$ processes, eventually coupling single input modes to thousands of output ones. We reproduced several features of topological systems, going beyond the results obtained with previous platforms~\cite{Cardano2017, Geraldi19, Derrico20, Derrico20bloch,qmDiColandrea24}, also exploring larger numbers of steps across 1D and 2D models.
The step-by-step analysis of QW dynamics demonstrates the benefit of reconfigurability, setting this platform apart from earlier static implementations.
The complexity of the explorable evolutions is only limited by the resolution of the devices. Deviations from theoretical predictions, which increase as expected with the number of modes, are mainly attributed to the reduced resolution and larger pixel size of the last SLM used in this experiment.

Our proof-of-concept demonstration lays the basis for future experiments which will benefit from a platform that is compact, programmable and is now ready-to-use. The proposed circuit technology targets translation-invariant operations with increasing complexity by keeping the amount of optical losses the same. Combined with reconfigurability, this unique feature will allow us to employ the same platform to explore multi-photon quantum protocols~\cite{paneru2024nonlocal}. 
\section{Methods}
\subsection{Quantum walks as space-dependent polarization transformations}
Quantum walks are realized as a sequence of space-dependent translations and coin rotations. With the definition provided in Eq.~\eqref{eq:tx}, these operators exhibit translational symmetry. In our encoding, the translations are implemented by LCMSs known as $g$-plates, acting as standard polarization gratings, while the coin rotations are implemented with standard waveplates~\cite{Derrico20}. $g$-plates are thin structured media whose optical action is described by a space-dependent Jones matrix:

\begin{equation}\label{eqn:gplate}
G_x(\delta)=\begin{pmatrix}
\cos(\delta/2) && i\sin(\delta/2)e^{-2i\pi x/\Lambda}\\
i\sin(\delta/2)e^{2i\pi x/\Lambda} && \cos(\delta/2)
\end{pmatrix}.
\end{equation}
If $\Lambda=2\pi/\Delta k_{\perp}$, a $g$-plate couples neighbouring modes as defined in Eq.~\eqref{eqn:opticalmodes}. Devices coupling modes along the $y$ direction have an analogous form. A quantum walk, either in 1D or 2D, is implemented by cascading multiple such devices to realize the required steps~\cite{Derrico20}. If diffraction can be neglected, the overall action of $N$ devices can be modeled as a complex, space-dependent transformation:

\begin{equation}
U=\iint \text{d}x\,\text{d}y \, \left( J_N \cdot J_{N-1} \cdot \dots \cdot J_1 \right) \otimes \ketbra{x,y},
\end{equation}
where $J_i$ represents the Jones matrix of the $i$-th plate (with the $xy$ dependence omitted for brevity). Defining $\mathcal{U}(x,y) = J_N \cdot J_{N-1} \cdot \dots \cdot J_1$, one obtains the expression in Eq.~\eqref{eqn:digonal}.

\subsection{Experimental setup}
\begin{figure}
    \includegraphics[width=\linewidth,  trim = 0.25cm 1.7cm 8.1cm 0.25cm, clip]{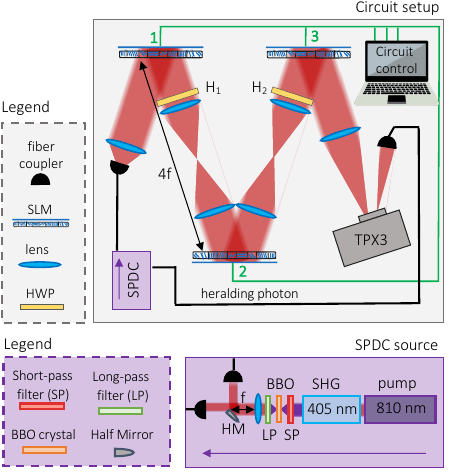}
    \caption{\textbf{Experimental Setup.} The setup consists of three SLMs arranged in a relay imaging system and simultaneously controlled by a LABVIEW software (Circuit control). The second SLM is sandwiched between two HWPs, $H_1(22.5 \text{\textdegree})$ and $H_2(-22.5 \text{\textdegree})$.
    A two-photon source based on SPDC produces pairs of temporally correlated photons, with one photon routed to the circuit and the other detected directly by the TPX3CAM camera to herald the potential arrival of its partner. The source is based on a Ti:Sa pulsed beam at $810$~nm (pulse duration $120$~fs, repetition rate $1$~GHz, output power $2$~W), which is frequency-doubled via Second Harmonic Generation (SHG) in a nonlinear crystal, yielding laser light at $405$~nm with $200$~mW average power. After SHG, the beam is focused (f = $400$~mm) onto a $\beta$-Barium Borate (BBO) crystal cut for type-I phase matching. Idler and signal photons are separated using a half mirror (HM) placed in the far field of the crystal. Experiments with classical light are carried out by replacing the SPDC source with a spatially filtered $810$~nm diode laser.}
    \label{fig:experimental setup}
\end{figure}

A heralded single-photon source was used to perform a QW experiment in the single-photon regime. Strongly correlated photon pairs were generated via SPDC, and coupled to single-mode fibers, with one photon serving as the herald and the other undergoing the QW. The SPDC source is described in Fig.~\ref{fig:experimental setup}. The setup consists of three LCOS-SLMs by Hamamatsu arranged in a relay imaging system. SLM-1 and SLM-2 belong to the X13138 series, with $ 1272\times 1024$ pixels and a $12.5$ $\mu$m pixel pitch, while SLM-3 belong to the X10468 series and has a lower number of pixels, $792 \times 600$, and a wider pixel pitch of $20$ $\mu$m.  As shown in Fig.~\ref{fig:experimental setup}, two 4-$f$ systems with equal focal-length lenses (200 mm) image SLM-2 onto SLM-1 and SLM-3 onto SLM-2, ensuring all three SLMs lie in the same plane within reasonable error. The incidence angle on SLM-2 and SLM-3 is below $9$\textdegree, to ensure approximately normal incidence. SLM-2 is sandwiched between two standard HWPs at $\pm 22.5$\textdegree. 
By post-selecting photons arriving simultaneously with the heralding ones at the camera, we reconstruct the distribution of the photons through the QW. All the classical experiments are carried out by replacing the SPDC source with a $810$ nm diode laser source.

A crucial step in the experiment is aligning the $3$ holograms displayed on the SLMs. The second and third SLMs are initially aligned with respect to the first by matching the center of their holograms to that of the first SLM in the image plane. However, this method introduces some uncertainty about the exact center positions. To refine the alignment, a laser is sent through the setup, and the positions of the hologram centers are fine-tuned by optimizing the resulting QW distributions. An additional calibration step is performed to fine-tune the phase discretization (gray level corresponding to the phase $2\pi$) of each SLM by optimizing the first diffraction order.\\

\subsection{Automatic control}
We developed a control software in order to synchronise the SLMs and data acquisition using LabVIEW 2021. Each SLM acts as a monitor and is controlled over HDMI, displaying grayscale holograms with bit depth values ranging from $0$ to $255$ representing phase changes of $0$ to $2 \pi$. The displayed holograms can be repositioned or replaced in real time, with a camera placed in the focal plane of the Fourier lens after the third SLM used for data acquisition. This camera's feed is displayed live in the control panel to allow for real-time analysis and control of the experiment.

In order to calculate the similarity of the distributions for 1D and 2D QWs as described in Eq.~\eqref{similarity}, the central walker site is located and placed within a square on the camera feed. For each step in the QW, the corresponding holograms are displayed, followed by the addition of new walker spots to accommodate the increased number of steps. Each walker site is assigned a probability corresponding to the total intensity at that walker site divided by the total intensity within all walker sites. This is then the experimentally measured probability distribution.

To achieve the reconstruction of the MCD outlined in Sec.~\ref{subsubsec:QM}, the centroid of the intensity distribution at the camera is captured and recorded for each of the $21 \times 21$ positions of the holograms. The $x$ and $y$ values of the centroid are recorded for each measurement.

\subsection{Analytical solution for the holograms}\label{subsec:analytical}
Here we provide the analytical solutions to Eq.~\eqref{eqn:solve}.
In our optical encoding, the mode mixing $U$ can be conveniently visualized as a space-dependent polarization transformation~\cite{DiColandrea2023,ammendola2024large}:
\begin{equation}
U=\iint \text{d}x\,\text{d}y\,\,\mathcal{U}(x,y)\ketbra{x,y},
\end{equation}
with $\mathcal{U}$ an SU(2) operator: 
\begin{equation}
\mathcal{U}(x,y)=\cos E\left(x,y\right)\sigma_0-i\sin E\left(x,y\right)\textbf{n}\left( x,y\right)\cdot\bm{\sigma},
\end{equation}
where ${E}$ is a real parameter and ${\textbf{n}=\left(n_1,n_2,n_3\right)}$ is a unit vector. 
The optical sequence of three SLMs and two HWPs, ${\mathcal{S}(x,y)=e^{i\frac{\delta_0}{2}}S_{0}(\delta_3)H_2S_{0}(\delta_2)H_1S_{0}(\delta_1)}$, is analogously decomposed as

\begin{equation}
\begin{aligned}
    \mathcal{S}(x,y) &= \textit{s}_0(x,y)\sigma_0 \\
    &\quad - i\left(\textit{s}_1(x,y)\sigma_1 + \textit{s}_2(x,y)\sigma_2 + \textit{s}_3(x,y)\sigma_3\right),
\end{aligned}
\end{equation}
where 
\begin{equation}
\begin{split}
\textit{s}_0 &=\sin\alpha\sin\beta,\\
\textit{s}_1 &=\cos\beta\sin\alpha,\\
\textit{s}_2 &= -\cos\alpha\sin\gamma,\\
\textit{s}_3 &= \cos\alpha\cos\gamma,
\end{split}
\end{equation}
with ${\alpha=\delta_2/2}$, ${\beta=\frac{\delta_1+\delta_3}{2}}$, and ${\gamma=\frac{\delta_1-\delta_3}{2}}$. The dependence on $(x,y)$ is omitted for ease of notation. Imposing ${\mathcal{U}=\mathcal{S}}$ at each transverse position yields
\begin{equation}\label{system}
\begin{split}
\sin\alpha\sin\beta&=\cos E,\\
\cos\beta\sin\alpha&=\sin E \sin\theta\cos\phi,\\
-\cos\alpha\sin\gamma&=\sin E\sin\theta\sin\phi,\\
\cos\alpha\cos\gamma&=\sin E\cos\theta,
\end{split}
\end{equation}
where we used the spherical parametrization of the vector $\textbf{n}$: ${n_1=\sin\theta\cos\phi}$, ${n_2=\sin\theta\sin\phi}$, and ${n_3=\cos\theta}$.
Four sets of solutions are found:
\begin{equation}
\begin{cases}
\alpha_1=\text{atan2}\left(\alpha_x,\alpha_y\right)\\
\beta_1=\text{atan2}\left(\beta_x,\beta_y\right)\\
\gamma_1=\text{atan2}\left(\gamma_x,\gamma_y\right)
\end{cases}
\end{equation}

\begin{equation}
\begin{cases}
\alpha_2=\text{atan2}\left(\alpha_x,\alpha_y\right)\\
\beta_2=\text{atan2}\left(\beta_x,-\beta_y\right)\\
\gamma_2=\text{atan2}\left(-\gamma_x,-\gamma_y\right)
\end{cases}
\end{equation}

\begin{equation}
\begin{cases}
\alpha_3=\text{atan2}\left(\alpha_x \sqrt{\frac{h}{h-1}},-\alpha_y\right)\\
\beta_3=\text{atan2}\left(-\beta_x,\beta_y\right)\\
\gamma_3=\text{atan2}\left(\gamma_x,\gamma_y\right)
\end{cases}
\end{equation}

\begin{equation}
\begin{cases}
\alpha_4=\text{atan2}\left(-\alpha_x,-\alpha_y\right)\\
\beta_4=\text{atan2}\left(-\beta_x,-\beta_y\right)\\
\gamma_4=\text{atan2}\left(-\gamma_x,-\gamma_y\right)
\end{cases}
\end{equation}
where atan2(${x,y}$) is the two-argument arctangent function, which distinguishes between diametrically opposite directions, and
\begin{equation}
\begin{split}
h&=\cos{E}^2-\sin{E}^2\sin{\theta}^2\cos{\phi}^2,\\
\alpha_x&=\cos\theta\sin E/\sqrt{1-h},\\
\alpha_y&=-\sin E \sin\theta\sin\phi/\sqrt{1-h},\\
\beta_x&=\sqrt{1-h},\\
\beta_y&=-\sqrt{h},\\
\gamma_x&=-\cos\phi \sin E \sin\theta/\sqrt{h},\\
\gamma_y&=-\cos E/\sqrt{h}.
\end{split}
\end{equation}
From the expressions for $\alpha$, $\beta$, and $\gamma$, the modulations for the SLMs' holograms $\delta_1$, $\delta_2$, and $\delta_3$ can be extracted. These are defined Mod$(4\pi)$, while holograms are physically defined up to $2\pi$. Since $S_0(\delta+2\pi)=-S_0(\delta)$, we use their value Mod$(2\pi)$, adding a minus sign when ${\delta_i(x,y) > 2\pi}$: $-S_0(\delta_i(x,y))$.
Figure~\ref{fig:fig1}(b) shows one among the possible sets of solutions $\delta_i(x,y)$ extracted for a 5-step 2D QW (protocol $U_2$, input state $\ket{H}$) that could be indistinctly used in the experiment.

\newpage

\bibliography{3SLMs}
\subsection*{Acknowledgement}
The authors thank Giulia Salatino for fruitful discussions and help in the preliminary stages of the experiment. 
\subsection*{Funding}
This work was supported by the PNRR MUR project PE0000023-NQSTI and PNRR
MUR project CN 00000013—ICSC, Canada Research Chairs (CRC), and Quantum Enhanced Sensing and Imaging (QuEnSI) Alliance Consortia Quantum grant. MGA further acknowledges support from MITACS and the Italy-Canada Innovation Award.
\subsection*{Competing interest}
The authors declare that they have no competing interests.
\subsection*{Data and materials availability}
All data are available in the main text or the supplementary materials.

\clearpage
\onecolumngrid
\renewcommand{\figurename}{\textbf{Figure}}
\setcounter{figure}{0} \renewcommand{\thefigure}{\textbf{S{\arabic{figure}}}}
\setcounter{table}{0} \renewcommand{\thetable}{S\arabic{table}}
\setcounter{section}{0} \renewcommand{\thesection}{S\arabic{section}}
\setcounter{equation}{0} \renewcommand{\theequation}{S\arabic{equation}}
\onecolumngrid

\begin{center}
{\Large Supplementary Material for: \\Compact and programmable large-scale optical processor in free space}
\end{center}
\vspace{1 EM}

\section*{Supplementary Data}

\section{Time-resolved 2D QW$\text{s}$}
\begin{figure}
\includegraphics[width=\linewidth,  trim = 0.25cm 0.75cm 0.25cm 0.75cm, clip]{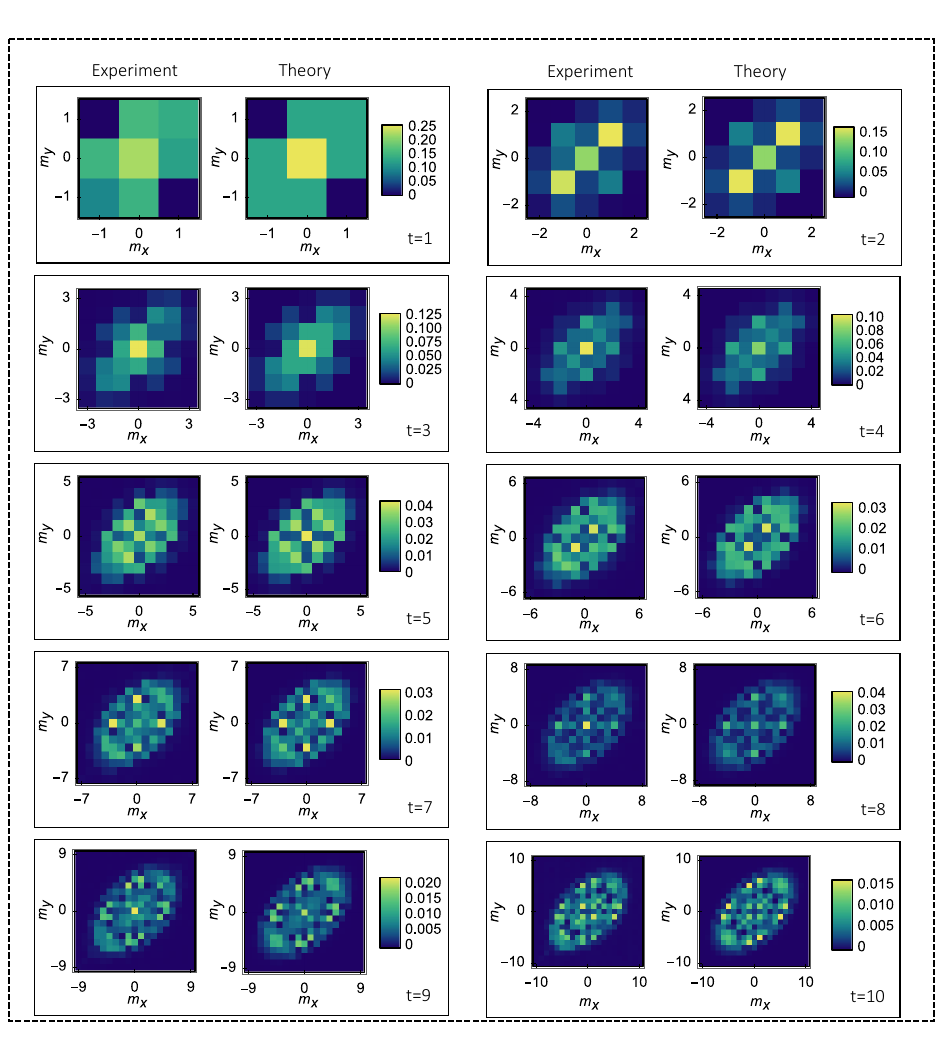}
    \caption{\textbf{2D QWs.} Experimental and theoretical distributions for each time step of the 2D QW protocol $U_1$, with input state $\ket{H}$, from $t=1$ to $t=10$.}
    
    \label{fig:suppfig1}
\end{figure}

\begin{figure}
\includegraphics[width=\linewidth,  trim = 0.25cm 0.25cm 0.25cm 0.25cm, clip]{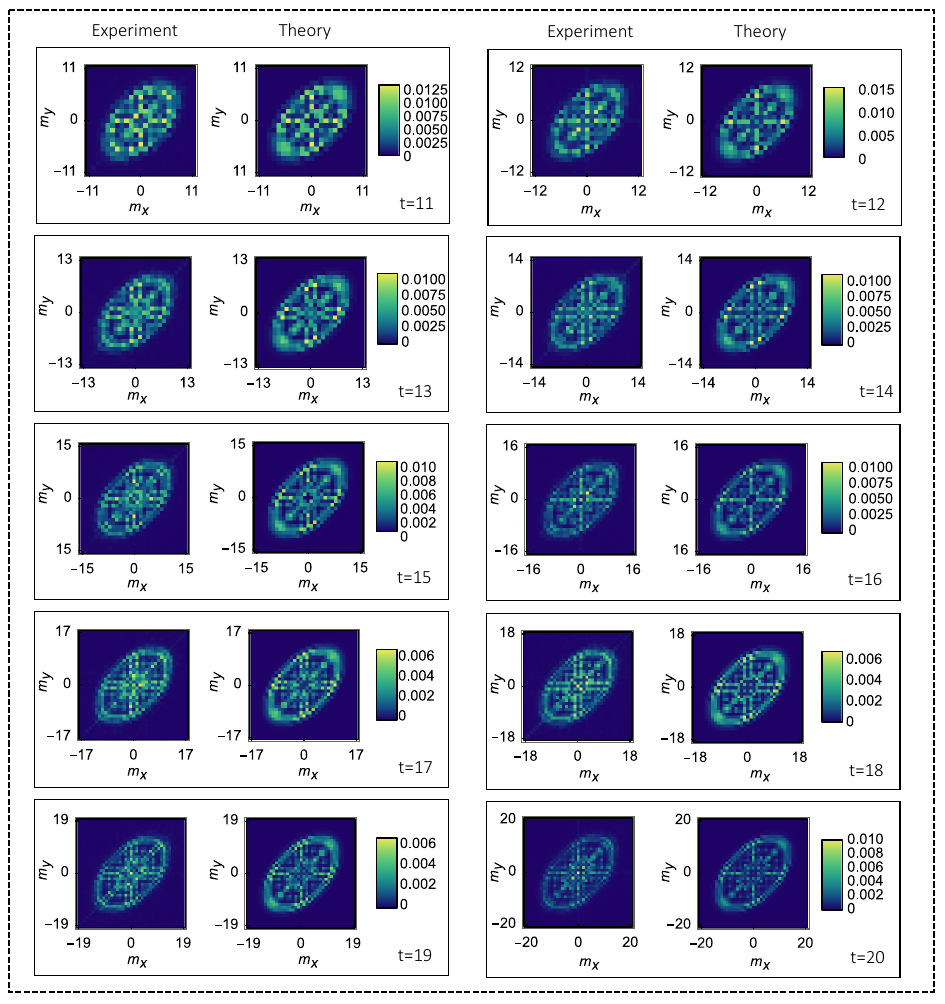}
    \caption{\textbf{2D QWs.} Experimental and theoretical distributions for each time step of the 2D QW protocol $U_1$, with input state $\ket{H}$, from $t=11$ to $t=20$.}
    
    \label{fig:suppfig2}
\end{figure}

\begin{figure}
\includegraphics[width=\linewidth,  trim = 0.25cm 0.25cm 0.25cm 0.25cm, clip]{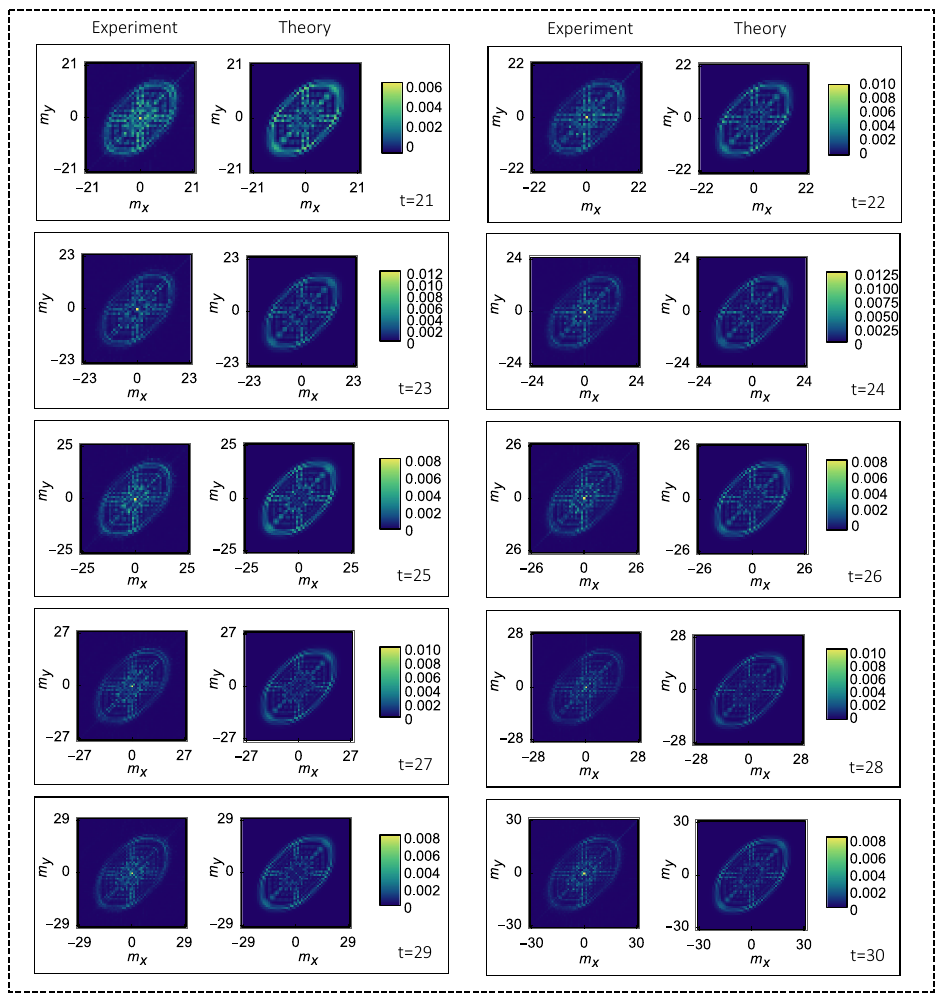}
    \caption{\textbf{2D QWs.} Experimental and theoretical distributions for each time step of the 2D QW protocol $U_1$, with input state $\ket{H}$, from $t=21$ to $t=30$.}
    
    \label{fig:suppfig3}
\end{figure}

\begin{figure}
\includegraphics[width=\linewidth,  trim = 0.25cm 0.25cm 0.25cm 0.25cm, clip]{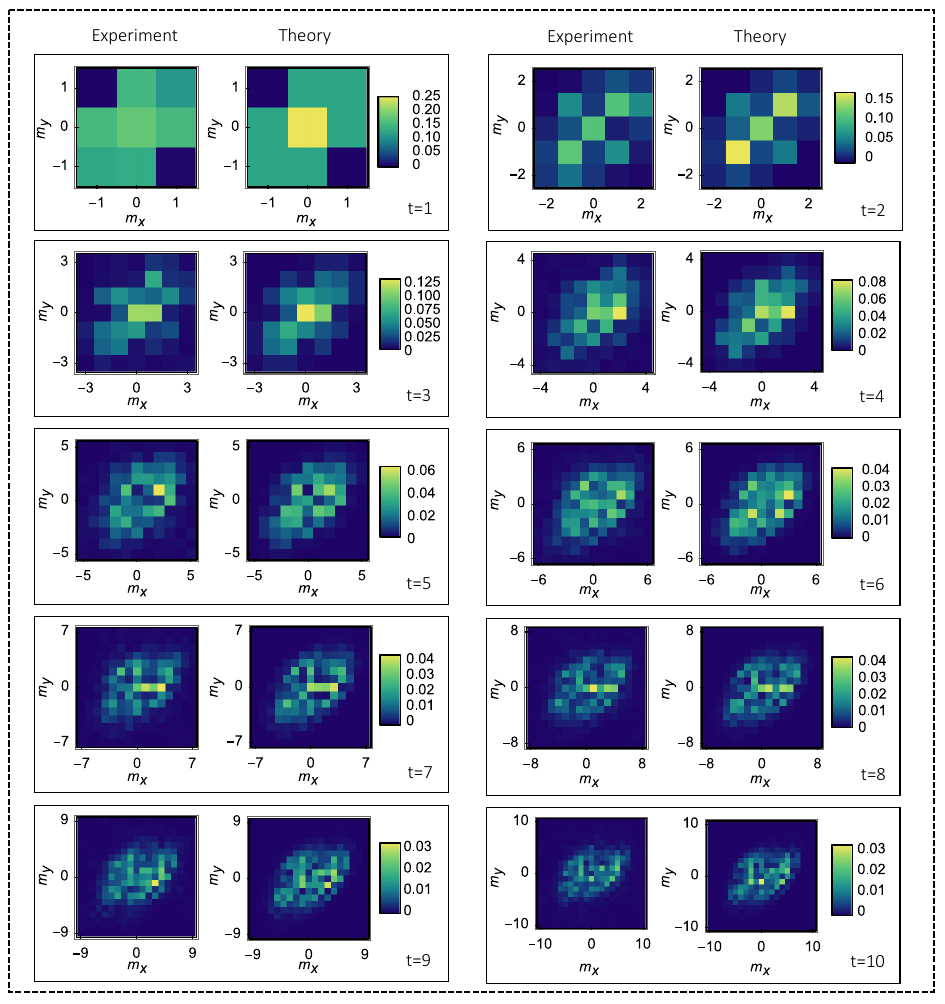}
    \caption{\textbf{Electric 2D QWs.}
    Experimental and theoretical distributions for a 2D QW (protocol $U_1$) with a constant electric field along the horizontal direction. We show each time step from $t=1$ to $t=10$, for an input state $\ket{H}$.}
    
    \label{fig:suppfig4}
\end{figure}

\begin{figure}
\includegraphics[width=\linewidth,  trim = 0.25cm 0.25cm 0.25cm 0.25cm, clip]{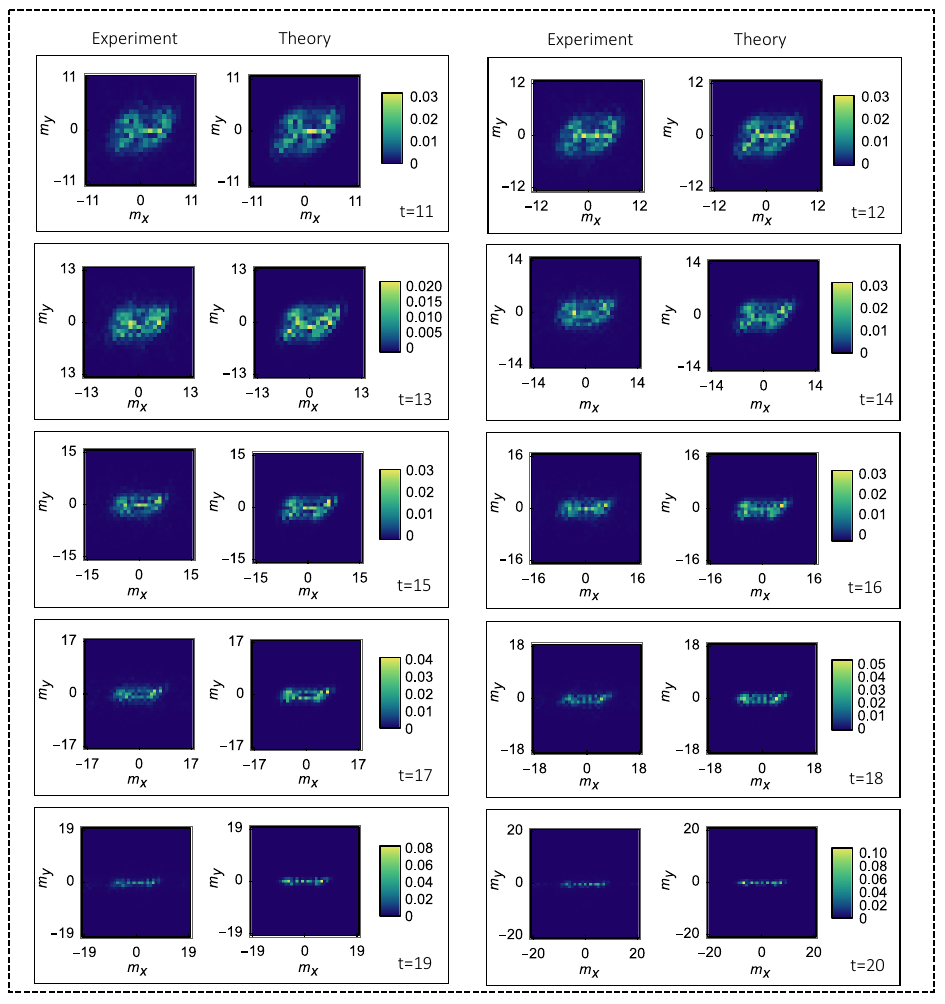}
    \caption{\textbf{Electric 2D QWs.}
  Experimental and theoretical distributions for a 2D QW (protocol $U_1$) with a constant electric field along the horizontal direction. We show each time step from $t=11$ to $t=20$, for an input state $\ket{H}$.}
    
    \label{fig:suppfig5}
\end{figure}

Figure~\ref{fig:suppfig1}-\ref{fig:suppfig5} show the results obtained for all the time steps for the 2D QW and the electric 2D QW introduced in the main text (see Fig.~\ref{fig:fig2}(b) and Fig.~\ref{fig:fig3}(c)).

\section{Superdiffusive transition in a 1D QW induced by temporal disorder}
\begin{figure}
\includegraphics[width=\linewidth,  trim = 0.3cm 0.5cm 0.4cm 0.25cm, clip]{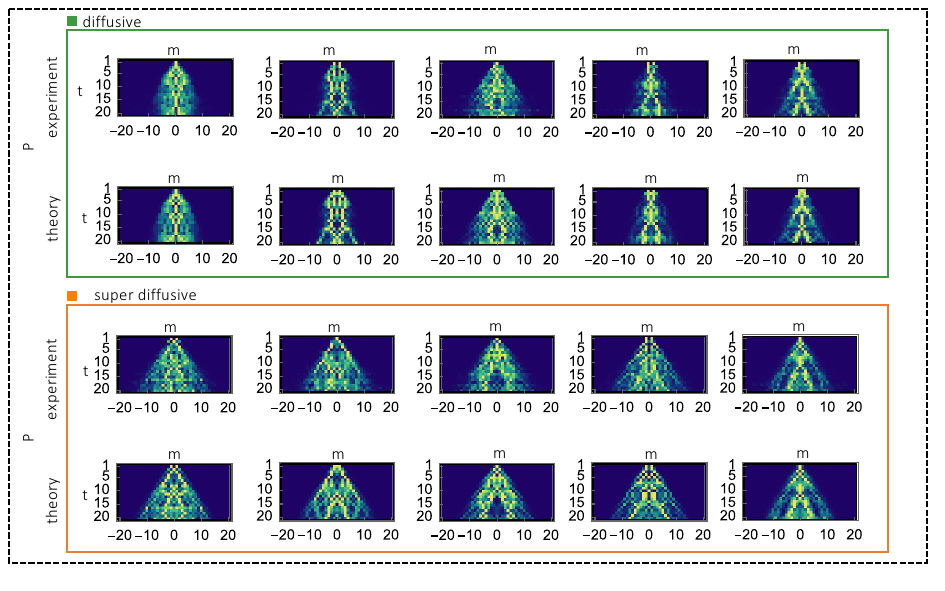}
    \caption{\textbf{Dynamical disorder in 1D QWs.}
    Experimental reconstruction (top) and theoretical prediction (bottom) of the output probability distribution $P$ for $20$ steps of a 1D QW with temporal disorder in five diffusive (green) and superdiffusive (orange) regimes.}
    
    \label{fig:suppfig6}
\end{figure}

Figure~\ref{fig:suppfig6} shows the results obtained for the different realizations of the disordered 1D QW experiment in the superdiffusive ($\Delta= 37.5\%$) and the diffusive ($\Delta= 87.5\%$) regime. These $5$ realizations have been used to compute the average value of the $\sigma^2(t)$ in the two cases, as shown in Fig.~\ref{fig:fig3}(b).

\section{Measurement of the Quantum Metric in a 2D chiral lattice system}
\begin{figure}
\includegraphics[width=\linewidth,  trim = 0.25cm 6.2cm 0.3cm 0.2cm, clip]{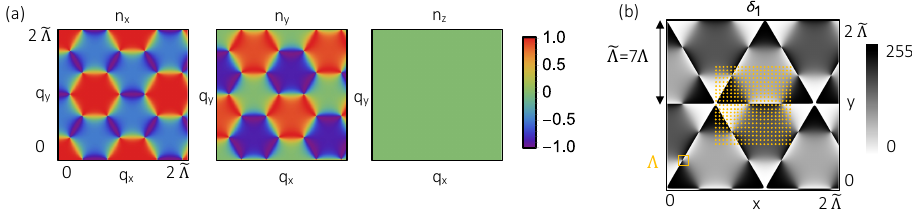}
    \caption{\textbf{Measurement of the quantum metric in a 2D chiral lattice system.} (a)~Eigenstructure $n_x(q_x,q_y)$, $n_y(q_x,q_y)$, $n_z(q_x,q_y)$ of the simulated flat-band graphene-like Hamiltonian.
     (b)~First hologram ($\delta_1(x,y)$) used for the MCD measurement described in Sec.~\ref{subsubsec:QM}. By magnifying the BZ with a 7$\times$ zoom factor, a wavefunction localized in the quasi-momentum space can be simulated. An automatic software shifts the three holograms simultaneously to simulate ${21 \times 21}$ different $\vec{q}_0$ values (yellow dots).} 
    \label{fig:suppfig7}
\end{figure}

The target evolution simulated in Sec.~\ref{subsubsec:QM} is generated from a flat-band graphene-like Hamiltonian featuring chiral symmetry, with chiral operator ${\Gamma=\sigma_z}$, which takes the following Bloch diagonal form:
    \begin{equation}
    \begin{aligned}
    \mathcal{H}_g(q_x,q_y)&= -\sigma_x [ \cos{((\sqrt{3} q_y -q_x)/2)}\\
   &+\cos{((\sqrt{3} q_y + q_x)/2)}+\cos{q_x}] + \\
    &\sigma_y [ \sin{((\sqrt{3} q_y -q_x)/2)}+\\
    &-\sin{((\sqrt{3} q_y + q_x)/2)}+\sin{q_x}].
    \end{aligned}
    \end{equation}
Figure~\ref{fig:suppfig7}(a) shows its eigenstates structure $n_i(q_x,q_y)$, where $i \in \{x,y,z\}$. Figure~\ref{fig:suppfig7}(b) shows one of the three holograms ($\delta_1$) used to measure the MCD, as described in Sec.~\ref{subsubsec:QM}. We set the size of the magnified BZ to $\Tilde{\Lambda}=7 \Lambda$. The Labview routine controlling the setup (described in Methods) simultaneously shifts the three holograms and centers the beam in $21 \times 21$ different $\vec{q}_0$ values, corresponding to the real positions ${(x_0,y_0)=(q_{0x},q_{0y})\Tilde{\Lambda}/2\pi}$ (yellow dots).

\end{document}